# R&D Towards Cryogenic Optical Links

**Mark Christiansen[a], Raphael Galea[b,1], Datao Gong[c], Suen Hou[e], David Lissauer[d], Chonghan Liu[c*], Tiankuan Liu[c], Veljko Radeka[f], Pavel Rehak[f,2], John Sondericker[c], Ryszard Stroynowski[c], Da-Shung Su[e], Peter Takacs[f], Helio Takai[d], Valeri Tcherniatine[d], Ping-Kun Teng[e], Craig Thorn[d], Annie C. Xiang[c], Jingbo Ye[c], Bo Yu[f]**

[a] *Department of Electrical Engineering, Southern Methodist University, Dallas TX 75275, U.S.A.*
[b] *Department of Physics, Columbia University, New York, NY 10027, U.S.A.*
[c] *Department of Physics, Southern Methodist University, Dallas TX 75275, U.S.A.*
[d] *Physics Department, Brookhaven National Laboratory, Upton, NY 11973, U.S.A.*
[e] *Institute of Physics, Academia Sinica, Nangang 11529, Taipei, Taiwan*
[f] *Instrumentation Division, Brookhaven National Laboratory, Upton, NY 11973, U.S.A.*
[1] *Now at National Research Council of Canada, Ottawa, ON, K1A0R6, Canada*
[2] *Deceased*

*E-mail*: kentl@smu.edu

ABSTRACT: A number of critical active and passive components of optical links have been tested at 77 K or lower temperatures, demonstrating potential development of optical links operating inside the liquid argon time projection chamber (LArTPC) detector cryostat. A ring oscillator, individual MOSFETs, and a high speed 16:1 serializer fabricated in a commercial 0.25-μm silicon-on-sapphire CMOS technology continued to function from room temperature to 4.2 K, 15 K, and 77 K respectively. Three types of laser diodes lase from room temperature to 77 K. Optical fibers and optical connectors exhibited minute attenuation changes from room temperature to 77 K.

## Contents

## 1. Introduction

The liquid argon time projection chamber (LArTPC) is a detector proposed for the long baseline neutrino experiment [1]. Operating optical links inside a liquid argon cryostat has the advantages of low thermal load, high data bandwidth, low noise, and low power consumption. In this paper we present electrical and optical measurement results on critical components of optical links from room temperature to 77 K (liquid nitrogen temperature) or 4.2 K (liquid helium temperature). We have identified a CMOS technology and several passive components and found promising laser diodes for the development of optical links operating inside the liquid argon cryostat.

## 2. Test setup

### 2.1 Circuits based on a SoS CMOS technology

Individual transistors [2], a ring oscillator and a serializer [3] tested were fabricated in a commercial 0.25 µm silicon-on-sapphire (SoS) CMOS technology. All the NMOS and PMOS transistors tested were laid out with a large W/L ratio by an elongated layout. The transistors were 40 µm wide and 0.25 µm long. The characteristic $I_D$-$V_{GS}$ curves were measured using a picoammeter and three programmable DC power supplies. A reed relay switch array controlled by a USB digital-input-output card was used to connect the terminals of the transistors to the picoammeter and the DC power supplies.

A 30-stage differential ring oscillator was implemented on the same die as the transistors which were tested. The output signals of the ring oscillator were transmitted through a twisted pair of cables and terminated with a 100-Ω resistor at the far end. The output waveform of the ring oscillator was recorded periodically using a real time oscilloscope and a differential probe.

A 5-Gbps 16:1 serializer was fed with 16 bit parallel data generated in a field-programmable gate array (FPGA). The serial output, which was a pseudorandom binary sequence (PRBS), of the serializer was recorded using a real time oscilloscope and a differential probe.

### 2.2 Optical fibers and connectors

A single mode (SM) optical fiber and a multi mode (MM) optical fiber have been tested. Both fibers were 50 meters long. However, the length of the part immersed in the liquid nitrogen was only 48 meters long.

Two types of LC connectors (mating sleeves), one with Zirconia sleeve for single mode and the other with Phosphor Bronze sleeve for multi mode, have been tested from room temperature to 77K. For each type five connectors were linked in series with fibers which had been tested, so when we calculated insertion loss changes of the connectors at low temperature the performance of the fibers were excluded. In both tests we used commercial small form-factor pluggable (SFP+) optical transmitters as the laser sources and a light wave multimeter to measure the optical power going through the fibers and connectors.

### 2.3 Laser diodes

A vertical cavity surface emitting laser (VCSEL) diode, a distributed feedback (DFB) laser diode and a Fabry Perot (FP) laser diode have been tested. The L-I-V characteristic curves of each laser diode were measured using a digital multimeter, a programmable DC power supply and a light wave multimeter. The optical spectra of each laser diode were measured using an optical spectrum analyzer (OSA).

### 2.4 Liquid helium setup

The measurements were performed parasitically with the regular operation of the Electron Bubble Chamber (EBC) [4]. A cryostat which operates between temperatures of 1 K and 300 K was designed to cool the EBC. The device under test (DUT) was located in the vacuum between the EBC and a liquid helium reservoir. The temperature of the DUT was determined by the temperatures of the surrounding heat shields. The temperature was monitored through a silicon diode clipped near the DUT.

The transistors and ring oscillators were tested using this setup. For all the components, the temperature cycle was performed twice. The resistors were kept working at 15 K for one hour at the first temperature cycle and 11 K for one hour at the second temperature cycle respectively. The ring oscillator was kept working at 4.2 K for one hour.

### 2.5 Liquid nitrogen setup

The DUT was attached at the bottom of a stepping motor system which was controlled via a computer. A Dewar which contained liquid nitrogen was put on the floor just below the bottom of the stepping motor system. The temperature depended on the distances between the DUT and the liquid nitrogen surface. The temperature of the DUT was monitored by using a silicon diode attached near the DUT.

The serializers, optical fibers, connectors and laser diodes were tested using this setup. The stepping motor system made the temperature cycles, include the temperature varied from room temperature to 77K and the temperature varied from 77K to room temperature. Each component was temperature cycled at least twice and kept working at 77K more than one hour.

## 3. Test results

### 3.1 The circuits based on the SoS CMOS technology

#### 3.1.1 Individual transistors

The characteristic $I_D$-$V_{GS}$ curves of NMOS and PMOS transistors are shown in the Figure 1.The curves becomes steeper as the $V_{DS}$ increases from 0.1 V to 1.2 V. Transconductance ($g_m$) and threshold voltage ($|V_T|$) were extracted from the $I_D$-$V_{GS}$ curves with $|V_{DS}|$ = 0.1 V at the linear range.

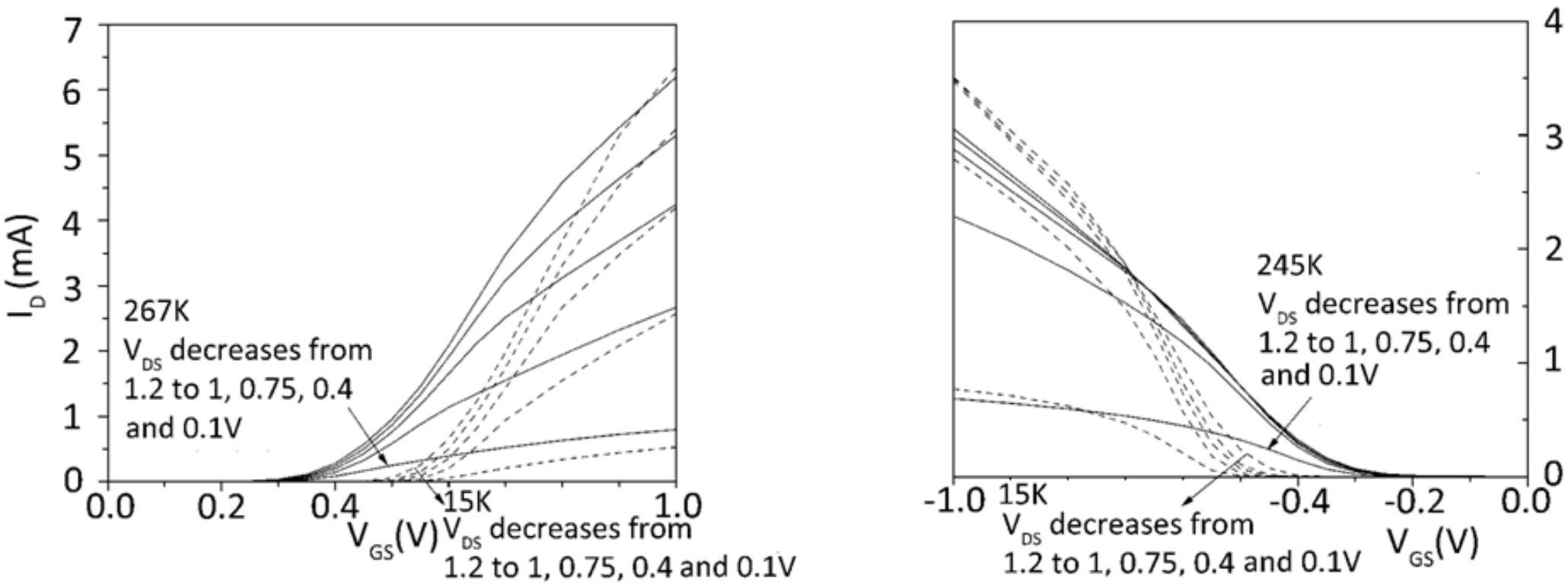

Figure 1. the $I_D$ vs. $V_{GS}$ at different temperatures
Left : NMOS transistor, right : PMOS transistor

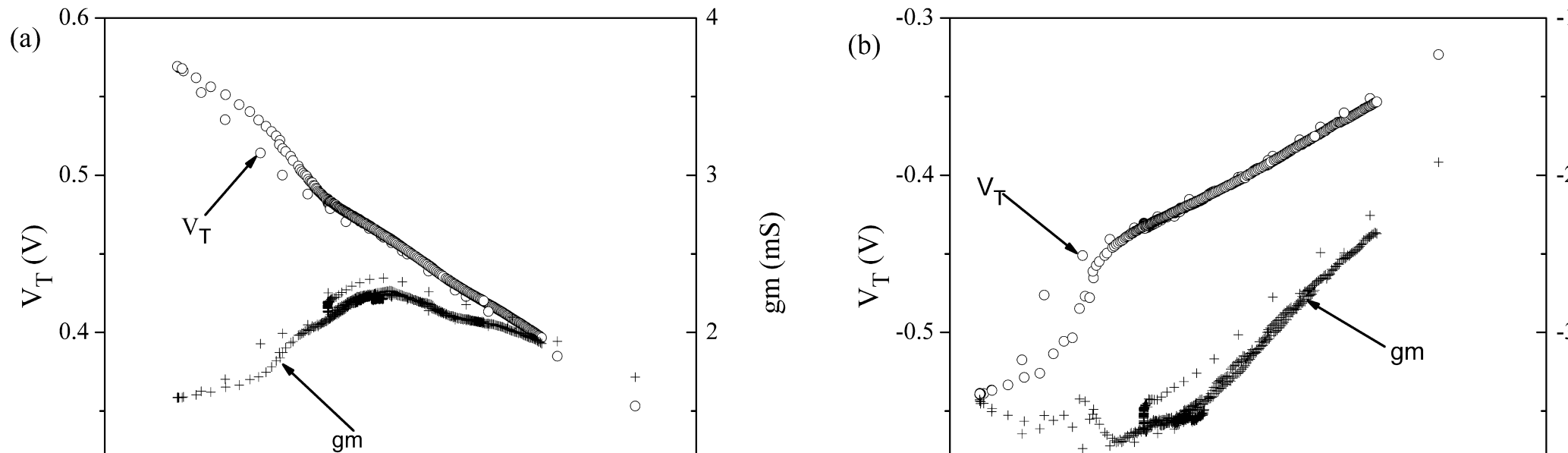

Figure 2. $|V_T|$ vs. temperature and $g_m$ vs. temperature
Left : NMOS transistor, right : PMOS transistor

It is shown in Figure 2 that the absolute values of the threshold voltages increase linearly above 77 K when the temperature decreases, consistent with the common transistor model [5]. The threshold voltage is not only inversely proportional to the temperature, but also has a complex relationship with Fermi energy which is also affected by temperature. The Fermi energy is roughly a constant above a certain temperature which dependents on doping concentration and is 77 K here, so the threshold voltage increases just linearly as the temperature decreases. Below this certain temperature the Fermi energy is not a constant anymore and decreases as the temperature increases. The variation of Fermi energy below the certain temperature breaks the linear relationship between the threshold voltage and temperature.

The absolute values of transconductance increase at first, then decrease when the temperature decreases. This is caused by the temperature dependence of the mobility [6]. The two scattering mechanisms that influence carrier mobility are lattice scattering which causes an increase in mobility with decreasing temperature and impurity scattering which causes a decrease in mobility with decreasing temperature. The impurity scattering dominates at low temperature whereas the lattice scattering dominates at relatively high temperature. As we know, the $g_m$ is proportional to the carrier mobility. Therefore the temperature dependence of the transconductance is as in Figure 2.

### 3.1.2 The ring oscillator

The waveform of the ring oscillator at 4.2 K is shown in Figure 3. The waveform has higher amplitude and faster transient time than those at room temperature. The oscillation frequency versus temperature relationship is shown in Figure 4. The frequency increases about 50% with the temperature decreasing from room temperature to 4.2 K.

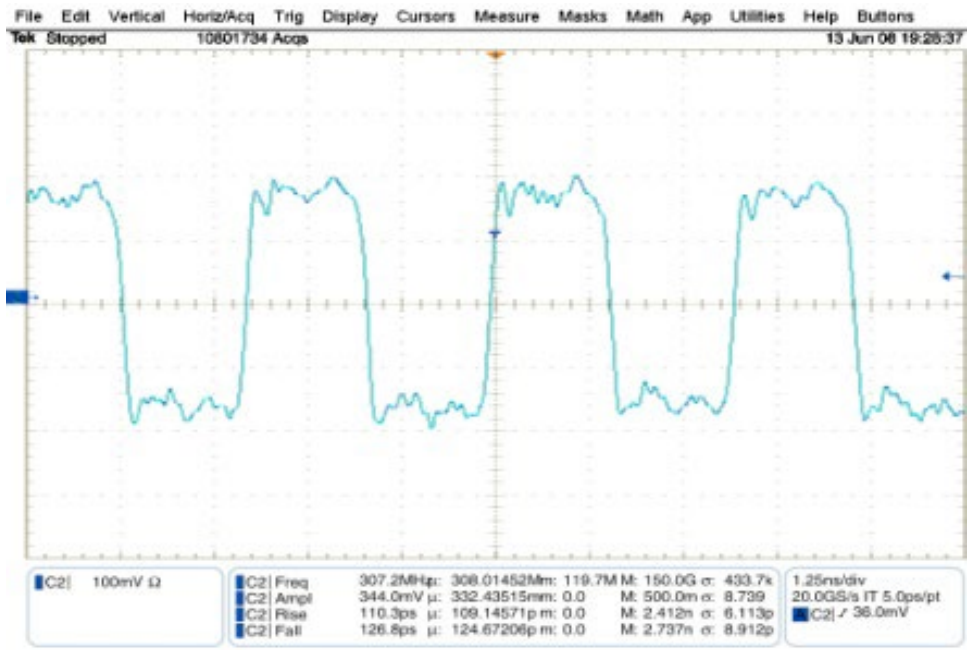

Figure 3. the ring oscillator waveform at 4.2K

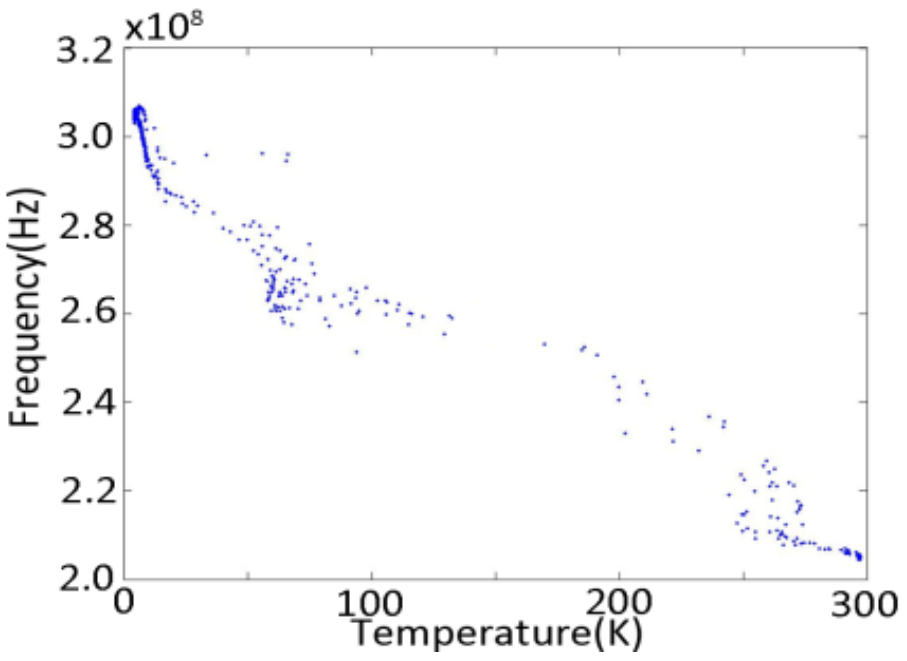

Figure 4. frequency versus temperature of the ring oscillator

### 3.1.3 The serializer

The eye diagrams of the serializer which was working at 5 Gbps are shown in Figure 5. The serializer at 77 K has wider eye opening, faster transient time, smaller jitter and larger amplitude than at room temperature. The mobility of the carrier increasing at low temperature causes the larger amplitude and faster transient time at low temperature. The smaller jitter is due to the smaller thermal noise at low temperature.

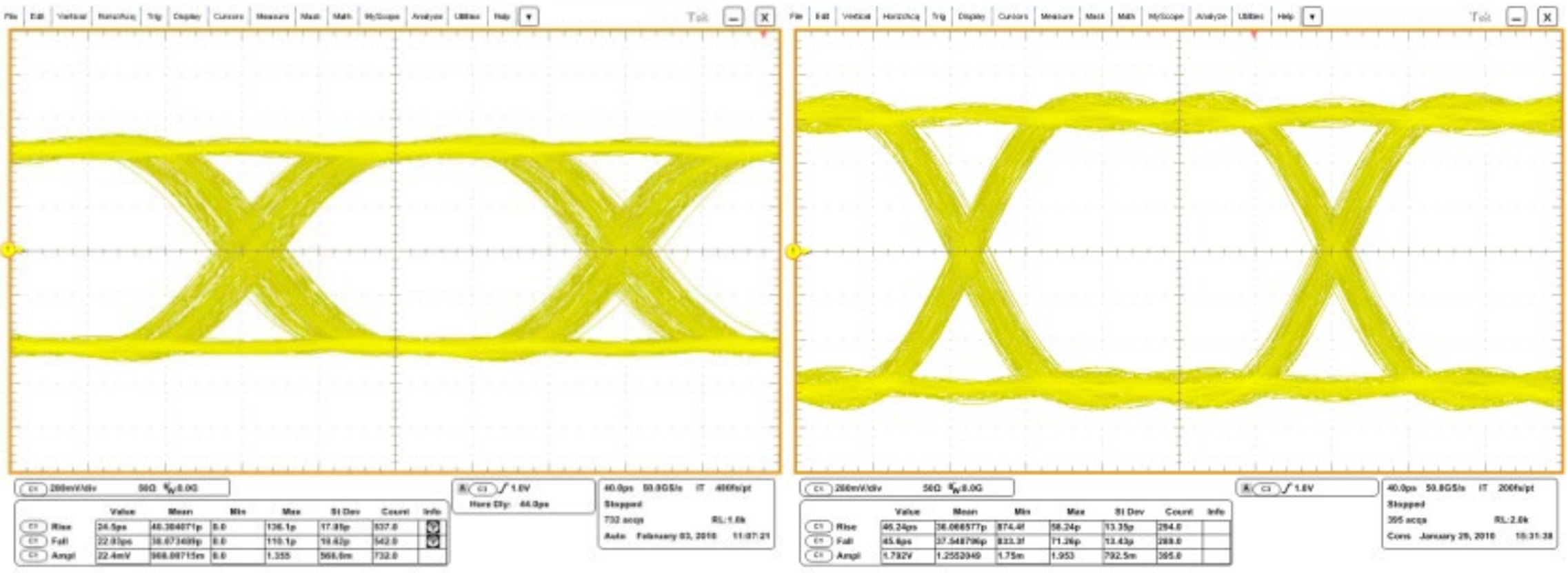

(a) room temperature (b) 77 K

Figure 5. eye diagrams of the serializer

### 3.2 Optical fibers and connectors

How the fiber optical power, which was transmitted by the fiber, changes with the temperature is shown in Figure 6. The cooling half cycle which is from room temperature to 77K and the warming half cycle which is from 77K to room temperature form one full temperature cycle. The two half cycles are not overlapped, because the DUT is much larger than the temperature sensor and the temperature of the sensor changes faster than the DUT.

The optical power, which was transmitted by the connectors linked with fibers, changed with temperatures is shown in Figure 7. As mentioned in section 2.2, we calculated the insertion loss at 77K excluding the performance of the fiber at 77K.

The optical power attenuation increases by 0.034 ± 0.004 ± 0.015 dB/m and 0.005 ± 0.004 ± 0.002 dB/m for multi mode and single mode fibers respectively. The insertion loss changes with temperature is 0.139 ± 0.011 ± 0.020 dB per connector and -0.284 ± 0.013 ± 0.014 dB per connector for multi mode and single mode connectors respectively. Each error expression consists of a system error and a measurement error. Given a usual optical power budget in a link system to be around 10 dB, the small power loss in the fibers and connectors at 77 K can be easily accommodated.

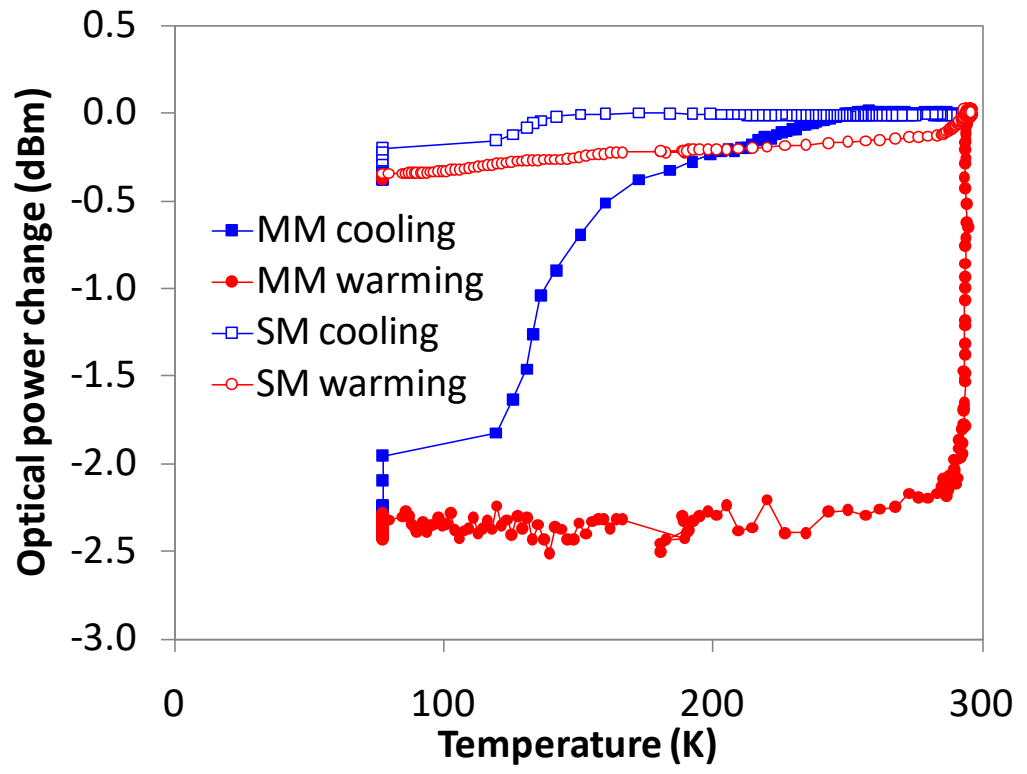

Figure 6. the optical power vs. temperature

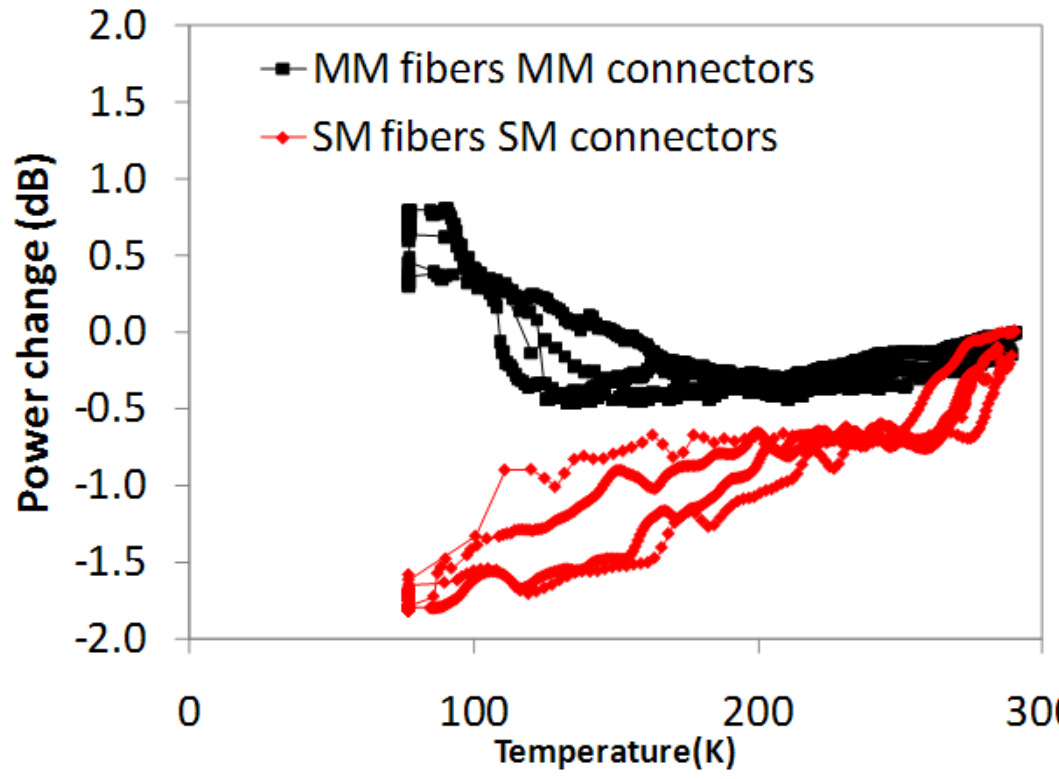

Figure 7. power changes vs. temperature

### 3.3 Laser diodes

The characteristic I-V, L-I curves and optical spectra of three laser diodes are shown in Figures 8-10. The threshold voltages of all three laser diodes increase from room temperature to 77 K. The threshold currents of DFB and FP laser diode decrease when temperature decreases from room temperature to 77 K, whereas that of the VCSEL increases.

The threshold currents of these three lasers behave differently as temperature decreases, because of their different structures. The FP laser does not have Brag reflector, the threshold current is only affected by the quantum well gain. The quantum well gain increases as temperature decreases, so the threshold current decreases as temperature decreases. The DFB laser and VCSEL laser have Brag reflectors, the minimum threshold current depends on how the gain spectrum aligns with the lasing mode which is determined by the reflectors and cavity. If they are aligned optimally at room temperature, the threshold current increases as temperature decreases, like the VCSEL tested here. On the other hand, if they are aligned optimally at low temperature, the threshold current decreases as temperature decreases, like the DFB laser tested here.

The light slope efficiencies of VCSEL and FP almost do not change at different temperatures, but the light slope efficiency of DFB drops when the temperature is close to 77 K. In optical spectra, when the temperature changes from room temperature to 77 K, the center wavelength shifts toward the short wavelengths and the spectral width becomes narrow.

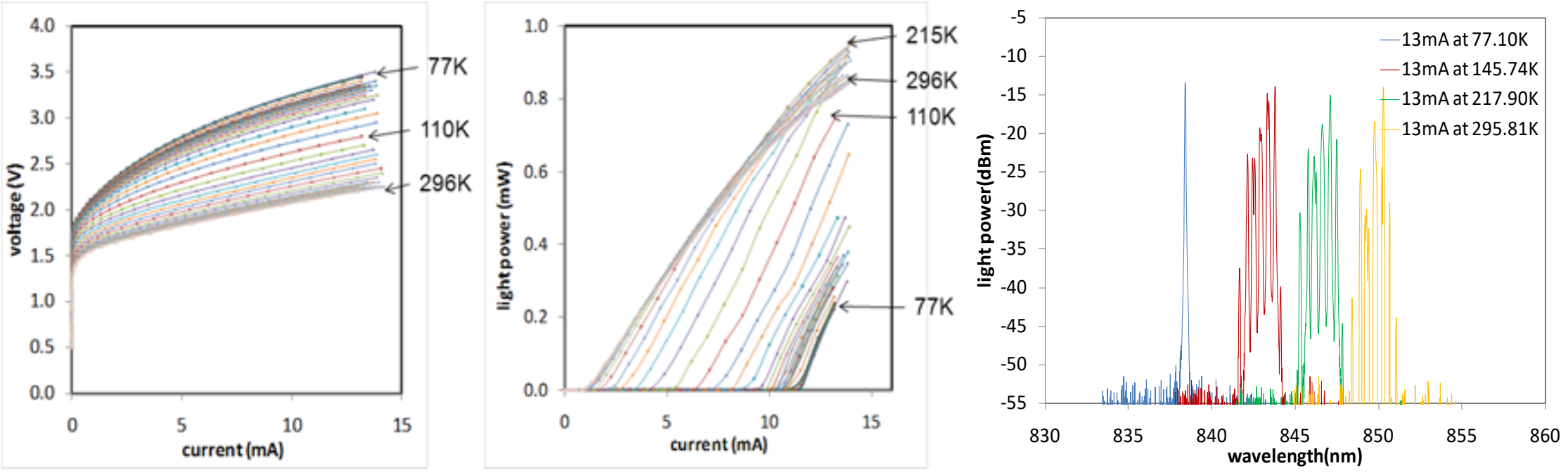

Figure 8. I-V curves, L-I curves and optical spectra of the VCSEL laser diode

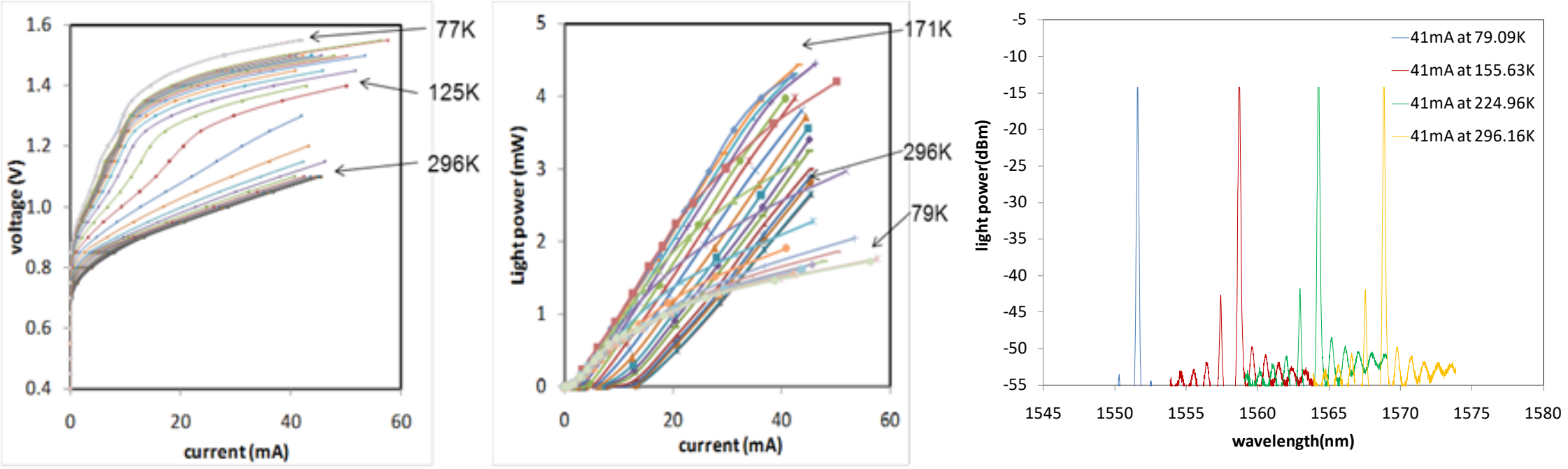

Figure 9. I-V curves, L-I curves and optical spectra of the DFB laser diode

DFB and FP laser diodes are the possible candidates used in optical links for liquid argon time projection chamber. The VCSEL lasers from other venders may need to be tested. We also need to study the reliabilities on these laser diodes.

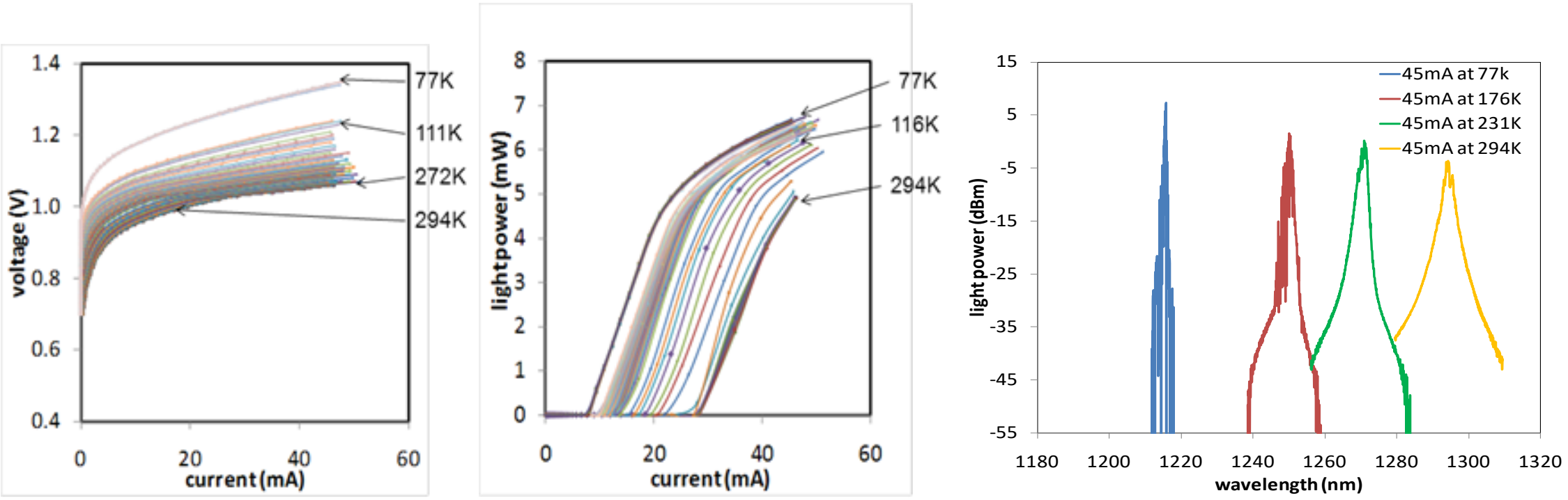

Figure 10. I-V curves, L-I curves and optical spectra of a FP laser diode

## 4. Conclusion

Ring oscillators, individual MOSFETs, and a 16:1 serializer fabricated in a commercial 0.25 µm SoS CMOS technology function from room temperature to 4.2 K, 15 K, and 77 K respectively. Optical fibers and optical connectors exhibit small attenuation changes and insertion loss changes respectively at 77 K. VCSEL, DFB and FB laser diodes lase throughout temperature range 300 K-77 K. The tests demonstrate the feasibility of optical links operating inside the liquid argon time projection chamber.

## Acknowledgments

The authors would like to express deepest appreciation to Gary Evans, Jim Guenter and Bob Biard for resources and technical consultation. The authors would like to thank Pritha Khurana, Arthur Mantie, Mark Schuckert and Tengyun Chen who provided us the laser diodes. The authors would also like to thank Todd Huffman, Anthony Weidberg, Francois Vasey and Jan Troska for beneficial discussion.